\documentclass[aps,pra,reprint,twocolumn,showpacs,showkeys,address,bibliography]{revtex4-2}

\usepackage{graphicx}
\usepackage{dcolumn}
\usepackage{bm}
\usepackage{braket}
\usepackage{xcolor}
\usepackage[normalem]{ulem}
\usepackage{appendix}
\usepackage[colorlinks,bookmarks=false,citecolor=blue,linkcolor=red,urlcolor=blue]{hyperref}
\usepackage{mathtools}
\usepackage[mathlines]{lineno}

\begin{document}

\preprint{APS/123-QED}

\title{Observation of effects of inter-atomic interaction on Autler-Townes splitting in
cold Rydberg atoms}

\author{Silpa B S$^1$}
\author{Shovan Kanti Barik$^1$}
\author{Varna Shenoy$^{1,2}$}
\author{Soham Chandak$^2$}
\author{Rejish Nath$^{2}$}
\author{Sanjukta Roy$^1$}
\email{sanjukta@rri.res.in}

\affiliation{$^1$ Raman Research Institute, C. V. Raman Avenue, Sadashivanagar, Bangalore-560080, India.}
\affiliation{$^2$Department of Physics, Indian Institute of Science Education and Research, Pune 411008 Maharashtra, India}

\date{\today}

\begin{abstract}

We demonstrate the effect of inter-atomic interaction in highly excited Rydberg atoms via Autler-Townes splitting in cold atoms. We measure the Autler-Townes (AT) splitting of the  $5S_{1/2}, F=2 \rightarrow 5P_{3/2}, F'=3$ transition of $^{87}Rb$ atoms arising due to the strong coupling of the transition via the cooling beams used for the magneto-optical trap (MOT). The AT splitting is probed using a weakly coupled transition from $5P_{3/2}, F'=3$ state to highly excited Rydberg states for a wide range of principal quantum numbers $(n=35-117)$. We observe the AT splitting via trap-loss spectroscopy in the MOT by scanning the probe frequency. We observe a drastic increase in the broadening of the AT splitting signal as a result of interaction-induced dephasing effect in cold Rydberg atoms for highly excited Rydberg states with principal quantum number $n > 100 $. We explain our observations using theoretical modelling and numerical simulations based on the Lindblad Master equation. We find a good agreement of the results of the numerical simulation with the experimental measurements.
\end{abstract}

\maketitle


\section{\label{sec:level1}Introduction}


Rydberg atoms have emerged as one of the most promising platforms for quantum technology, with applications in quantum information processing \cite{saffman2010quantum, paredes2014all, Henriet2020quantumcomputing,shao24}, quantum sensing \cite{adams2019rydberg,fancher2021rydberg} and quantum simulation of condensed matter systems \cite{browaeys2020many,weimer2010rydberg,morgado2021quantum}. It is due to their exaggerated atomic properties, and in particular, long lifetimes and strong Rydberg-Rydberg interactions (RRIs). These interactions lead to the well-known phenomenon of Rydberg blockade, which is a crucial mechanism used to realize quantum gates and explore quantum many-body physics.
\par
Autler-Townes (AT) or dynamic Stark splitting occurs in a multi-level atom when a strong resonant oscillating field couples two levels, which is then probed by a weak field that involves a transition to a third state \cite{autler1955stark}. The probe transition splits into two components separated by an energy of Rabi frequency associated with the strong resonant transition \cite{PhysRevA.91.053842}. Recent advances in the field have significantly expanded the scope of AT spectroscopy, particularly by using highly excited Rydberg atoms \cite{PhysRevA.68.053407,2018_Raithel}. Notably, AT spectroscopy has been employed to measure transition dipole moments \cite{Piotrowicz_2011}, carry out high-resolution spectroscopy \cite{2022_Arimondo}, conduct precise measurements of DC, microwave or radio-frequency electric field strengths \cite{2025_Hou, 10.1063/1.4963106,10.1063/1.4984201,photonics9040250,sedlacek12,2021_Holloway,PhysRevApplied.16.024008,2014_Raithel}, including polarization-insensitive electrometry \cite{cloutman2024polarization} and probing RRI induced dephasing effects \cite{2013_Jia,DeSalvo2016, 2021_Jia, 2022_Cao, saakyan2023rydberg}. The efficiency of AT-based Rydberg electrometry can be further improved using magnetic fields \cite{2023_OE,PhysRevA.109.L021702,10.1063/5.0233994}. The interaction-induced dephasing affects the AT splitting in different ways; for instance, it reduces the AT splitting and increases the width of the AT spectral lines. Though the above studies have been on cold Rydberg gases, the AT spectra in thermal Rydberg atoms have also been extensively studied in an external magnetic field, which ensures an enhanced Zeeman splitting \cite{barik2024doppler}.

In this paper, we present our experimental observations of AT splitting in the optical regime, using cold Rb atoms trapped in a magneto-optical trap (MOT) and probed via Rydberg states with high principal quantum numbers $(n=35-117)$. Our experimental setup is designed to probe the atoms with high precision, enabling us to detect subtle AT effects with greater sensitivity. While other studies on AT splitting in cold atoms \cite{2013_Jia,DeSalvo2016, 2021_Jia, 2022_Cao, saakyan2023rydberg} probed Rydberg states with $n < 72$, we investigate the effect of interaction-induced dephasing in AT splitting for highly excited Rydberg states with $n > 100$. We demonstrate a sudden drastic increase in the broadening of the AT splitting for Rydberg states with $n > 100$. 
The experimental results agree very well with the numerical results obtained by using the Lindblad Master equation, where we model the system as an effective single atom. The effect of RRIs are incorporated through a dephasing term and a shift in the cooling-beam detuning.


The paper is structured as follows. The experimental setup and procedure are described in Sec.~\ref{exptsetup}. The nature of trap-loss spectroscopy is discussed in Sec.~\ref{tlsp}. The master equation governing the effective three-level atom picture is given in Sec.~{\ref{model}}. The experimental measurements of the AT splitting as well as their comparison to theoretical modeling for various cooling beam and probe beam detuning, as well as a wide range of principal quantum number of the Rydberg state are described in Sec.~{\ref{results}}.
    
\section{Experimental setup}
\label{exptsetup}
\begin{figure*}[hbt]
\includegraphics[width=0.95\textwidth]{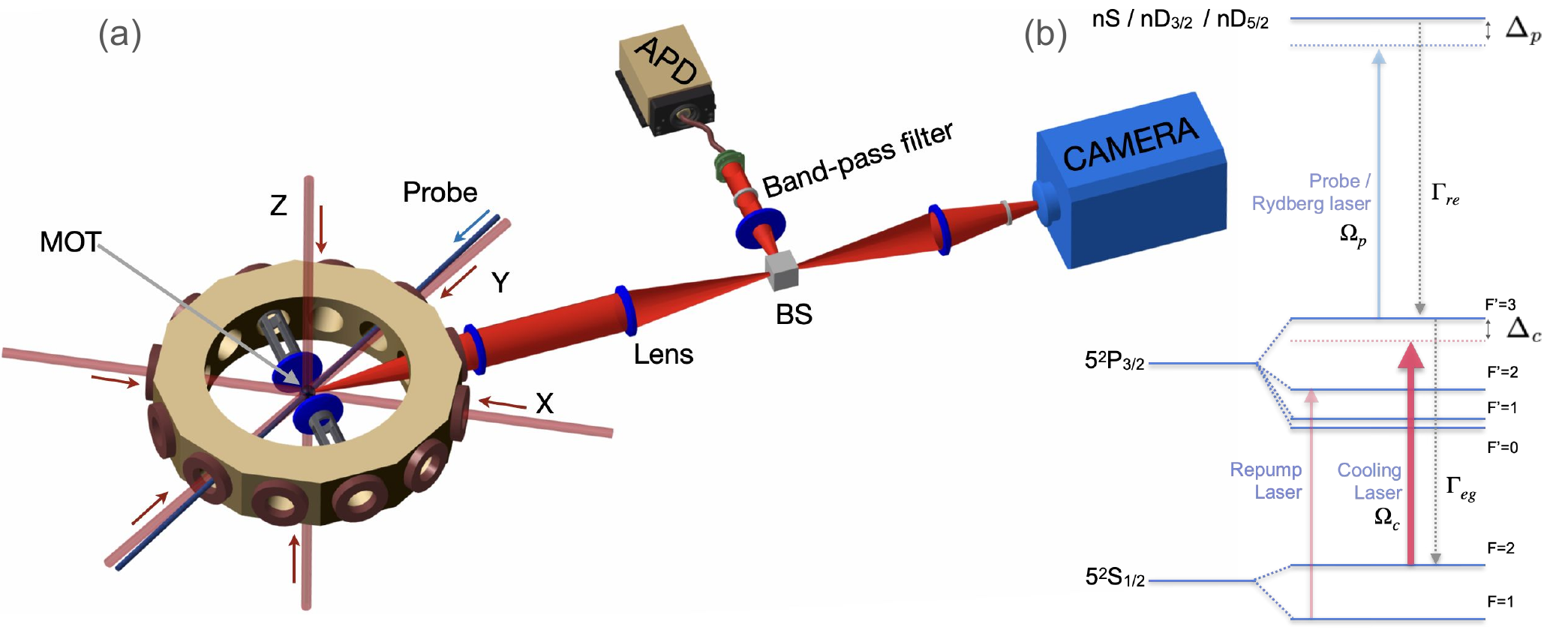}
\caption{\label{fig:wideExp}(a) The schematic of the experimental setup for Rydberg excitation of cold atoms in a \textsuperscript{87}Rb MOT and trap-loss spectroscopy. The fluorescence of the MOT is measured using an EMCCD camera and an Avalanche photo-diode (APD). (b) Energy-level scheme for two-photon Rydberg excitation along with the cooling and repumping laser beams of the MOT. The ground state and the intermediate state are coupled by the strong cooling beams of the MOT with a detuning of $\Delta_c$.}
\end{figure*}
We study the Autler-Townes splitting, involving Rydberg excitations, using trap-loss spectroscopy in cold \textsuperscript{87}Rb atoms in a MOT. The schematic diagram of the setup is shown in Fig.~\ref{fig:wideExp}(a) and the level scheme for laser cooling of \textsuperscript{87}Rb as well as Rydberg excitation is provided in Fig.~\ref{fig:wideExp}(b). In our experiment, the atoms are cooled and trapped in the MOT at a temperature of 200$\mu$K,  enabling high-resolution AT spectroscopy. The cooling beams of the MOT are red-detuned by $2\Gamma$ to the $5S_{1/2}, F=2\rightarrow 5P_{3/2}, F'=3$ transition. The cooling beams in the horizontal plane (X and Y beams in Fig~\ref{fig:wideExp}(a)) are not exactly perpendicular to one another due to the space constraints arising due to two high numerical aperture (HNA) lenses placed inside the vacuum chamber. The atoms are spatially confined in the MOT created using the cooling laser beams and a quadrupole magnetic field generated by two magnetic coils placed in an anti-Helmholtz configuration.

The laser, which is tuned to the transition $5S_{1/2}, F=1 \rightarrow 5P_{3/2}, F'=2$, repumps the atoms that are decayed into $5S_{1/2}, F=1$ ground state back to the cooling cycle. The cooling, as well as the repumping lasers, are derived from two DL100 Toptica lasers operating at 780 nm wavelength and frequency locked via a Lock-in regulator (LIR) using \textsuperscript{87}Rb saturation absorption spectroscopic signal as the frequency reference. The cooling laser is locked onto the spectroscopic signal corresponding to $5S_{1/2}, F=2 \rightarrow5P_{3/2}, F'=1-3$ cross-over and frequency shifted using an acousto-optical modulator (AOM) to make it red-detuned by $\sim 2\Gamma$ to $5S_{1/2}, F=2 \rightarrow 5P_{3/2}, F'=3$ transition. The parameter $\Gamma$ is the natural line width of the closed transition $5S_{1/2}, F=2 \rightarrow5P_{3/2}, F'=3$ of \textsuperscript{87}Rb valued $\sim 2\pi \times 6.065(9)$MHz. Similarly, the repumping laser is locked to the signal corresponding to the $5S_{1/2}, F=1 \rightarrow5P_{3/2}, F'=1-2$ crossover and frequency shifted using an AOM to the $5S_{1/2}, F=1 \rightarrow5P_{3/2}, F'=2$ transition. The total cooling power used for the trapping is $\sim 15 \; I_{sat}$, where $I_{sat}=3.58mW/cm^2$ is the saturation intensity for the given cooling transition. The steady-state cold atomic cloud in the MOT had a Gaussian diameter of $\sim 700\mu m$ and a peak atom number density of $5\times10^9$ atoms/$cm^{3}$. The temperature of atomic cloud measured using simplified time of flight fluorescence imaging method \cite{wang2023autler} is $\sim 200\mu K$. 

A laser beam of wavelength 480 nm,  is used to probe the strong cooling laser dressed two-level atomic transition in the cold atomic cloud in the MOT by coupling $5P_{3/2}, F'=3 \rightarrow nS/nD$ Rydberg levels as shown in Fig.~\ref{fig:wideExp}(b). This beam at 480 nm was derived from a Toptica TA-SHG pro frequency-doubled laser system with a seed laser frequency of 960 nm. With a Gaussian diameter of $\sim 1.5 mm$, this probe laser is aligned through the MOT, illuminating the cold atomic cloud. The two-photon Rydberg excitation of the cold \textsuperscript{87}Rb atoms are studied using the trap-loss spectroscopy technique \cite{cao2022dephasing,wang2023autler}. 
The Rydberg excited atoms leave the MOT cooling cycle and ballistically fly out of the trap \cite{saakyan2023excitation}. The fluorescence of the cold atomic cloud is recorded  using a single photon counting module (SPCM-AQ4C  from Excelitas) (Fig.~\ref{fig:wideExp}(a)), and the reduction in MOT fluorescence is directly proportional to the rate of Rydberg excitation. The trap-loss spectroscopy signal from the MOT fluorescence is recorded in each experimental run. The probe laser at 480 nm is scanned across an atomic transition from the intermediate state to one of the Rydberg states ($5P_{3/2}, F'=3 \rightarrow nS/nD$) \cite{2022_OC}. A wavelength meter (HighFinesse
WS8-2) is used to measure the frequency scanning rate of the probe laser, which is typically between 2 to 5 MHz per second and varied for different Rydberg levels according to the total line broadening of the transition. 

\begin{figure}
\includegraphics[width=0.47\textwidth]{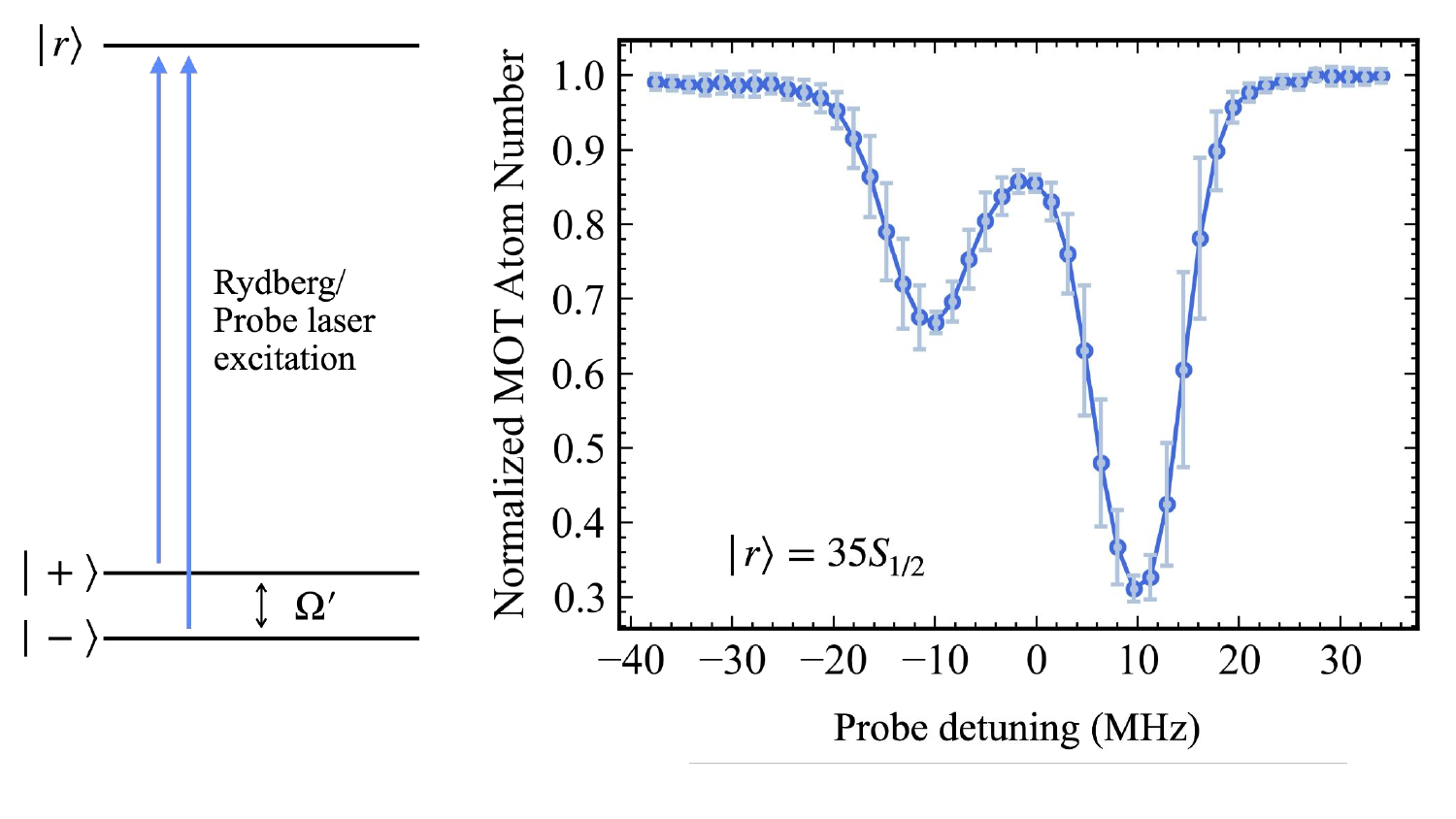}
\caption{\label{fig:RydTraploss} (Left) AT dressed split states where the probe couples the two dressed states to the Rydberg level $\ket{r}$. $\Omega'$ is the generalized Rabi frequency of the cooling transition. (Right) A typical trap-loss spectrum of \textsuperscript{87}Rb cold atoms in the MOT due to two-photon excitation to $\ket{r}$. The measurement has been carried out in a MOT with peak atom number density of $\approx 1 \times 10^{10} cm^{-3}$ by exciting the cold atoms to the Rydberg state $35S_{1/2}$. A total cooling beam intensity of $\approx 55 mW/cm^2$ and a probe beam intensity of $15 mW/cm^2$ was used in the measurement.}
\end{figure}

\begin{figure*}
\includegraphics[width=0.9\textwidth]{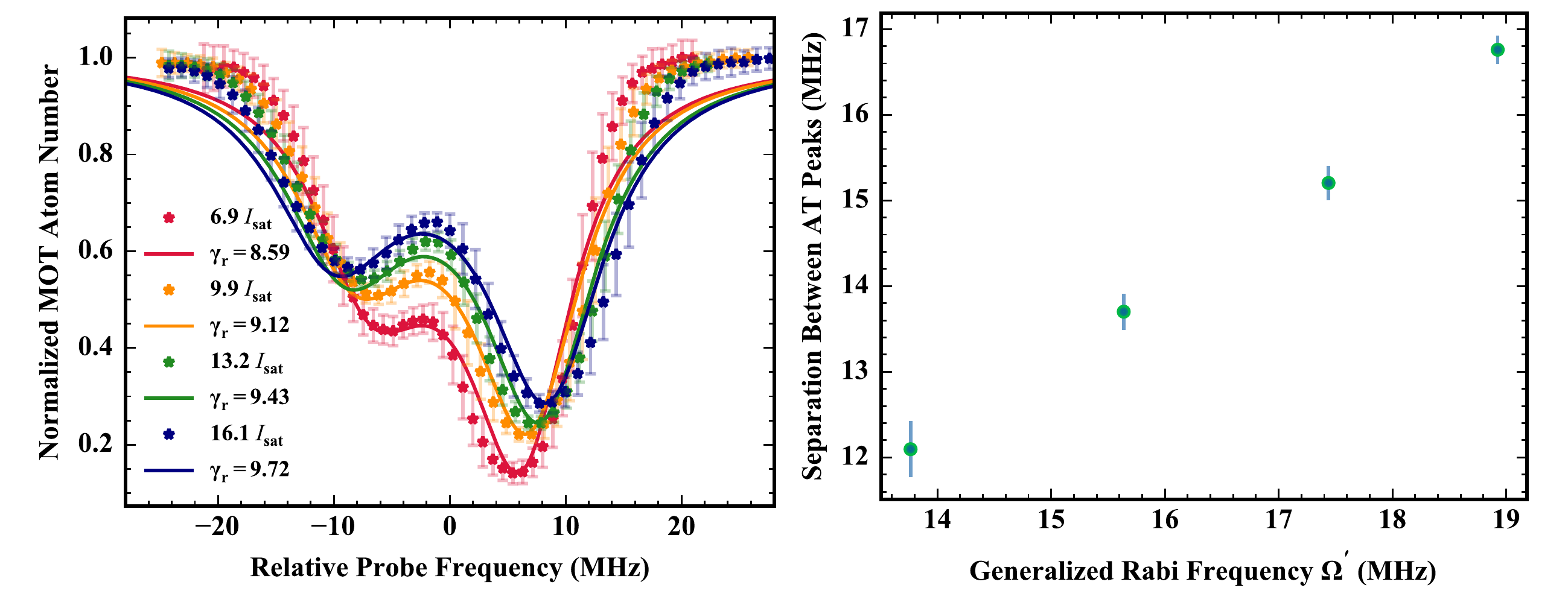}
\caption{\label{fig:35DiffCoolIntAT}(a) The Trap-loss spectra obtained for $\ket{r}=35S_{1/2}$ with different cooling beam intensities. The cooling beam intensities are given in the unit of saturation intensity for $5S_{1/2}, F=2 \rightarrow 5P_{3/2}, F'=3$ transition. The solid lines are the numerically estimated steady-state fraction of atoms left in the MOT ($1-f_R$) as a function of probe detuning $\Delta_p$ for corresponding experimental parameters. The parameter, interaction-induced dephasing $\gamma_r$ is deduced from the fit to the experimental data for different cooling Rabi frequencies $\Omega_c$. The cooling beam detuning $\Delta_c$ was optimized in the modelling using the experimental measurements.  The plot in (b) shows the amount of AT splitting vs the generalized Rabi frequency for cooling transition calculated with the corresponding cooling beam intensities. }
\end{figure*}

\section{Trap-loss spectroscopy of cold $^{87}{Rb}$ atoms}
\label{tlsp}
The amount of fluorescence signal collected from the cold atomic cloud is directly proportional to the total number of atoms $N$ trapped in the MOT in the $5S_{1/2}, F=2$ ground state. We can model the dynamics of the atom number $N(t)$ in the MOT having a constant loading rate $L$, using the rate equation, 
\begin{equation}
\frac{dN(t)}{dt}=L-\alpha N(t)
\label{dnt}
\end{equation}
 where $\alpha=\alpha_0 + \alpha_{exc}^{Ryd}(\omega_p)$ is the total single-atom loss rate.  $\alpha_0$ is the decay rate associated with the atom number loss from the MOT due to the collisions with the background thermal atoms inside the vacuum chamber and $\alpha_{exc}^{Ryd} (\omega_p)$ is the probe laser frequency ($\omega_p$) dependent decay due to the Rydberg excitation. The solution of Eq.~(\ref{dnt}) is of the form $N(t)=N_0(1-e^{-\alpha t})$ with $N_0 = L/\alpha$, considering no atoms are present in the MOT at $t=0$. In the absence of the probe beam, $N_0 = L/\alpha_0$. The steady-state atom number $N_0$ decreases as the probe laser frequency $\omega_p$ is tuned across an atomic transition to a Rydberg state. This decrease is proportional to the Rydberg excitation rate $\alpha_{exc}^{Ryd}(\omega_p)$.

To understand the nature of AT splitting, we model each atom as a three-level atom comprising of $\ket{g}\equiv |5S_{1/2}\rangle ($F=2$),\ket{e}\equiv|5P_{3/2}\rangle$ ($F=3$) and $\ket{r}$ corresponding to the ground, intermediate and the Rydberg states, respectively. The Hamiltonian of a three-level atom under the rotating wave approximation, in the $\{\ket{g},\ket{e},\ket{r}\}$ basis, is given by 
\begin{equation}
\label{h1}
    \hat H  = \frac{\hbar}{2}\begin{pmatrix}
    0 & \Omega_{c} & 0\\
\Omega_{c} & -2\Delta_{c} & \Omega_{p}\\
0 & \Omega_{p} & -2(\Delta_{c} + \Delta_{p})
\end{pmatrix},
\end{equation}
where $\Omega_{c}$ and $\Delta_c$ ($\Omega_{p}$ and $\Delta_p$) are the Rabi frequency and detuning of the control (probe) field coupling the $\ket{g}\rightarrow\ket{e}$ ($\ket{e}\rightarrow\ket{r}$) transition. In our experiments, we have $\Omega_{p} \ll \Omega_{c}$. Under the strong coupling, the two lowest states form the dressed states,
\begin{equation}
    \ket{\pm} = \frac{1}{\sqrt{2\Omega'}}\left(\pm\sqrt{\Omega'\pm\Delta_{c}}\ket{g} + \frac{\Omega_{c}}{\sqrt{\Omega'\pm\Delta_{c}}}\ket{e}\right)
\end{equation}
where $\Omega' = \sqrt{\Omega_{c}^2+\Delta_{c}^2}$ with energy eigenvalues $E_{\pm} = \frac{\hbar}{2}\left(-\Delta_{c}\pm\Omega'\right)$. Finally, we can write the Hamiltonian in the basis $\{\ket{-},\ket{+},\ket{r}\}$ as, 
\begin{equation}
 \hat H  = \frac{\hbar}{2}\begin{pmatrix}
-\Delta_{c}-\Omega' & 0 & \frac{\Omega_{c}\Omega_{p}}{\sqrt{2\Omega'^2-\Delta\Omega'}}\\
0 & -\Delta_{c}+\Omega' & \frac{\Omega_{c}\Omega_{p}}{\sqrt{2\Omega'^2+\Delta\Omega'}} \\
\frac{\Omega_{c}\Omega_{p}}{\sqrt{2\Omega'^2-\Delta\Omega'}} & \frac{\Omega_{c}\Omega_{p}}{\sqrt{2\Omega'^2+\Delta\Omega'}} & -2(\Delta_{c} + \Delta_{p})
\end{pmatrix}
\end{equation}
In Fig.~\ref{fig:RydTraploss}, the trap loss spectroscopy signal is shown as a function of the probe detuning ($\Delta_p$), which exhibits two minima in $N_0$ corresponding to the resonant Rydberg excitation from the dressed states $\left|{\pm}\right>$. The asymmetric depths of the minima \cite{2016_Ryabtsev} is due to the detuning $2\Gamma$ of the cooling beams of the MOT which is used as the control field giving rise to the AT splitting. The detuning of $2\Gamma$ facilitates Doppler cooling in the MOT required for realizing the cold atomic cloud.


The generalized Rabi frequency, $\Omega'= \sqrt{\Omega_c^2 + \Delta_c^2}$ provides the corresponding  Autler-Townes splitting. The cooling transition Rabi frequency is calculated as, $\Omega_c=E_c \mu_{ge}/\hbar$, with $\mu_{ge}$  is the dipole matrix element of the cooling transition and the total cooling field amplitude $E_c=\sqrt{2I_c/c\epsilon_0}$, where $c$ is the speed of light, $\epsilon_0$ is the vacuum permittivity and $I_c$ is the cooling beam intensity. The measurements are performed by probing the MOT atoms with a weak excitation to the various Rydberg $nS_{1/2}$ levels for a wide range of principal quantum numbers. The zero of the probe detuning in the plots is chosen as the midpoint of the AT doublet.

The trap-loss spectra are captured by scanning the frequency of a weak, unidirectional probe beam that illuminates the whole cold atomic cloud over the resonant Rydberg transition. The weak intensity of the probe beam means that the effects of radiation pressure which could potentially push the atoms out of the trap and the auto-ionization of Rydberg atoms are considered negligible. Nevertheless, if it exists, it can lead to a higher decay rate in MOT atom number, manifesting as a broadening in the AT spectra for higher probe intensity.




\section{Theoretical model: Master equation}
\label{model}

In the effective single three-level atom picture, we incorporate the Rydberg-Rydberg interactions as a dephasing effect \cite{Raitzsch2009}. Among the three levels, the intermediate excited state $\ket{e} = \ket{5P_{3/2}}$($F = 3$) has the highest decay rate. The dynamics of the system is governed by the Lindblad Master equation, 
\begin{equation}
    \frac{d\rho}{dt} = \frac{i}{\hbar}[\hat H,\rho] + \mathcal{L}[\rho],
    \label{me}
\end{equation}
where $\hat H$ is given in Eq.~(\ref{h1}) and $\mathcal{L}$ incorporates the decay from spontaneous emission and dephasing effects and is given by,
\begin{equation}
    \mathcal{L}[\rho] = \sum_{k}\gamma_{k}(L_{k}\rho L_{k}^{\dagger} - \frac{1}{2}L_{k}^{\dagger}L_{k}\rho -\frac{1}{2}\rho L_{k}^{\dagger}L_{k})
    \label{meLin}
\end{equation}
with $L_{k}$ being the jump operators \cite{Fleischhauer2005}.
\begin{equation}
    \mathcal{L} = \begin{pmatrix}
    \Gamma_{eg}\rho_{22} & -\frac{1}{2}\Gamma_{eg} \rho_{ge} & -\frac{1}{2}\gamma_3 \rho_{gr} \\
     -\frac{1}{2}\Gamma_{eg} \rho_{eg} & -\Gamma_{eg}\rho_{ee} + \Gamma_{re}\rho_{rr} & -\frac{1}{2}(\Gamma_{eg} + \gamma_3)\rho_{er} \\
     -\frac{1}{2}\gamma_3\rho_{rg} & -\frac{1}{2}(\Gamma_{eg} + \gamma_3)\rho_{re} & -\Gamma_{re}\rho_{rr}
    \end{pmatrix},
\end{equation}
where $\Gamma_{ij}$ is the spontaneous decay from the state $i$ to $j$. The decay rate $\gamma_3 = \gamma_r + \Gamma_{re}$ quantifies the dephasing of the Rydberg state, where, $\Gamma_{re}$ is the decay rate of the Rydberg state to the intermediate state, which is small, and $\gamma_r$ accounts for the dephasing due to the Rydberg-Rydberg interactions.  $\gamma_{r}$ is determined by an optimized fit of the numerical data to the experimental data (see Appendix.~\ref{aopt}) and is found to be several orders of magnitude larger than $\Gamma_{re}$ \cite{Zhang2014}. In our model, we use experimentally determined values for the probe and cooling Rabi frequencies, $\Omega_p$ and $\Omega_c$, and then optimize the interaction-induced dephasing rate, $\gamma_r$, and the cooling detuning, $\Delta_c$. Another effect of the interaction is to cause a change in the cooling beam detuning $\Delta_c$ \cite{DeSalvo2016}, 
which we then take as a free parameter, determined by optimizing the detuning with the experimental data. The decay rates, $\Gamma_{ij}$ are calculated using ARC calculator, which gives us $\Gamma_{eg} = 2\pi \times 6.01$ MHz. Considering the final measurements are taken after few milliseconds, we can assume the MOT atom numbers reached a steady state as discussed in Sec.~\ref{tlsp}. We solve Eq.~(\ref{me}) for the steady state, which provides us $\alpha(1-f_{R})$, where $f_R=\rho_{rr}$ and $\alpha$ are determined by scaling the blue-detuned peak from the experiment (the peak at the right in the AT splitting signal shown in Fig.~\ref{fig:RydTraploss}).


\section{Results}
\label{results}
\subsection{AT Spectra for various cooling and probe beam intensities}

\begin{figure*}
\includegraphics[width=0.9\textwidth]{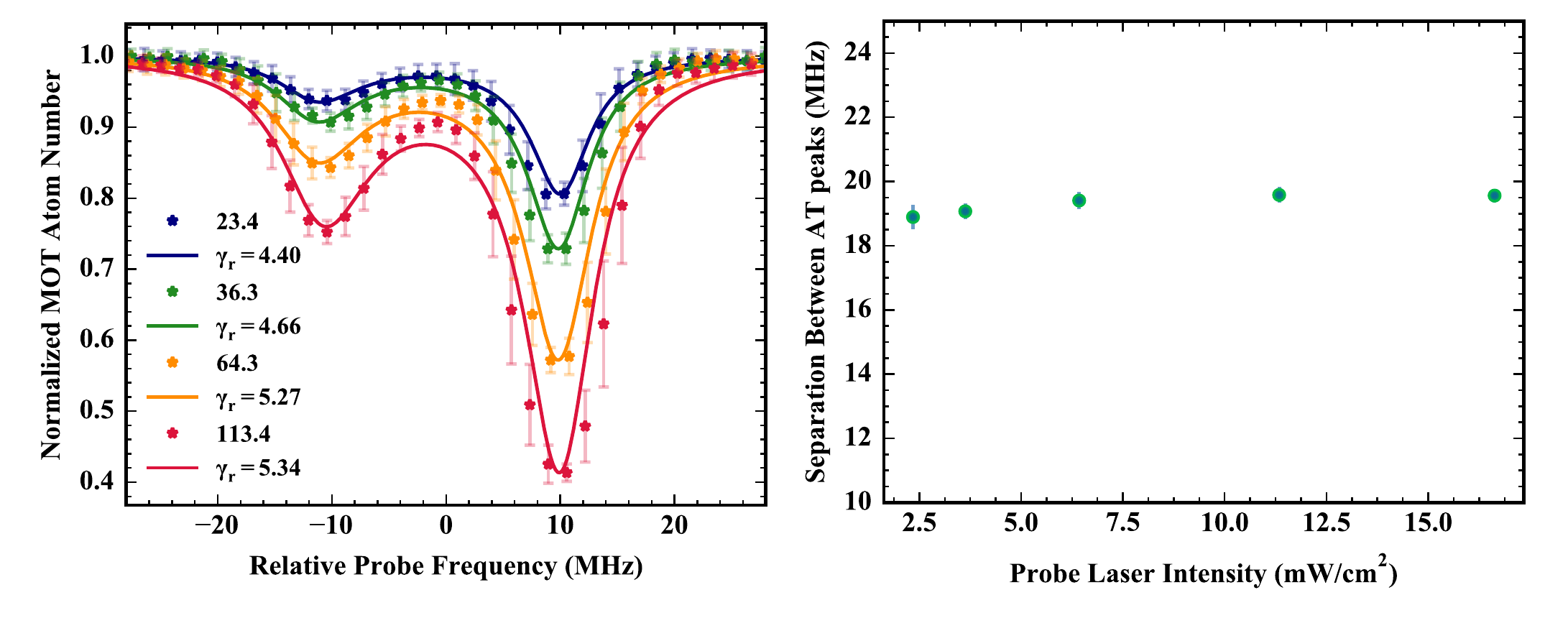}
\caption{\label{fig:35SdiffProbeIntAT}The Trap-loss spectra obtained for $\ket{r}=35S_{1/2}$ with different Rydberg/probe beam intensities measured in $W/m^2$ are shown in (a) The numerically estimated steady state fraction of atoms left in the MOT ($1-f_R$) as a function of probe detuning $\Delta_p$ with the shift in detuning and the interaction-induced dephasing fit to the experimental data for different probe Rabi frequencies $\Omega_p$ is shown as the solid line in (a) in comparison with the experimental results. (b) shows the amount AT splitting vs the Rydberg/probe beam intensities. } 
\end{figure*}


\begin{figure*}
\includegraphics[width=0.75\textwidth]{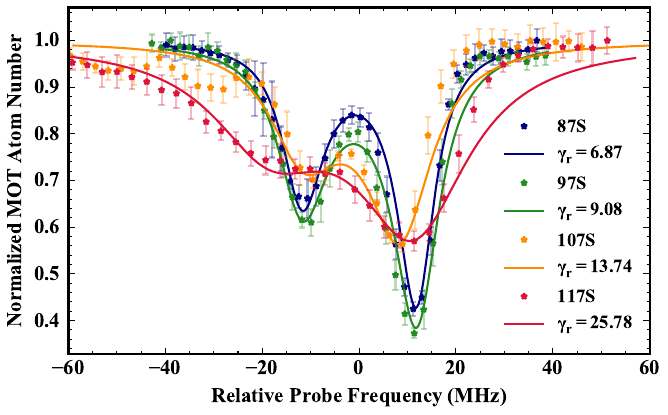}
\caption{\label{fig:RydIntATdiffN} The Trap-loss spectra obtained for different principal quantum number \textit{n} of \textsuperscript{87}Rb(Left) and the steady state fraction of atoms left in the MOT ($1-f_R$) for the corresponding \textit{n} levels obtained numerically (solid line). The measurements are carried out in a cold-atomic cloud of magneto-optical trap with cooling detuning of 10 MHz and a generalized Rabi frequency of 23.7 MHz. The probe laser intensity is chosen in such a way as to keep the $\Omega_p$ same for all the measurements($\approx$ 5.2 kHz).}
\end{figure*}

\begin{figure*}
\includegraphics[width=0.8\textwidth]{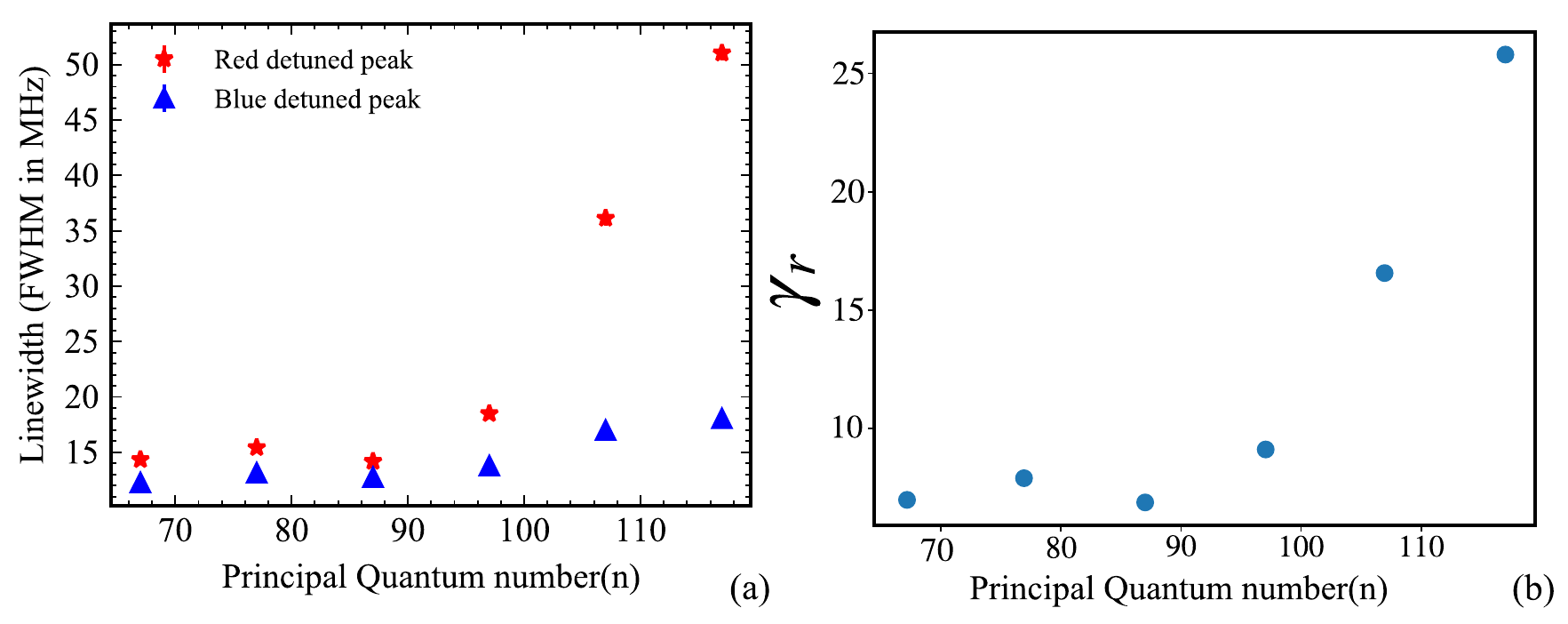}
\caption{\label{fig:RydIntlinewidthdiffN}The linewidth of the red and blue detuned dips obtained by curve fit using inverted Gaussian model plotted as a function of the principle quantum number \textit{n}(Left) The dephasing rate calculated for the different \textit{n} levels from fig. \ref{fig:RydIntATdiffN} (Right).}
\end{figure*}

Here, we report the trap-loss spectroscopy for various cooling and probe beam intensities for $\left|r\right> = 35S_{1/2}, F''=2$. The effect of varying the cooling beam intensity on the AT splitting is shown in Fig.~\ref{fig:35DiffCoolIntAT}(a) for $\Delta_c = 7.9$MHz  and a probe beam intensity of $32.5mW/cm^2$ (Expt). We observe that the AT splitting increases with the cooling beam intensity as expected. The generalized Rabi frequency calculated for the cooling transition vs the AT splitting measured from trap-loss spectroscopy for various cooling beam intensities is shown in Fig.~\ref{fig:35DiffCoolIntAT}(b),  and both quantities are expected to be equal. The mismatch between the two as shown in Fig.~\ref{fig:35DiffCoolIntAT}(b) can be due to an overestimation of the intensity of the cooling beam along the path. This means that the actual cooling beam intensity seen by the cold atomic cloud is lower than the estimated intensities using the direct optical power measurements. The solid lines in Fig. \ref{fig:35DiffCoolIntAT} (a) presents the numerically calculated steady-state fraction of atoms as a function of the probe detuning using the master equation in Eq.~(\ref{me}) and is found to be in excellent agreement with the experimental results. The numerical results show that increasing the cooling beam intensity leads to a higher $\gamma_r$, indicating an enhanced RRIs.


 In contrast with the above results of varying $I_c$, changing the probe beam intensity $I_b$ has negligible influence on the AT splitting, however exhibits larger atom loss with increasing intensity, as shown in Fig.~\ref{fig:35SdiffProbeIntAT}(a). The corresponding numerical results are shown in Fig.~\ref{fig:35SdiffProbeIntAT}(b). The numerical optimization results reveal that the dephasing rate $\gamma_3$ also increases with an increase in probe beam intensity, due to the formation of more Rydberg atoms at higher intensities. This is supported by the augmented atom number loss shown in Fig.~\ref{fig:35SdiffProbeIntAT}(a). The broadening in AT spectra with an increase in $I_b$ is attributed to power broadening and the rise in the dephasing rate $\gamma_3$.

The effect of $\gamma_r$ is to cause a broadening in the AT peaks \cite{Singer2004}, and we observe that $\gamma_r$ increases mildly on increasing $I_p$. The power broadening can also be a major contributor to the line broadening with the increasing probe beam intensity.

\subsection{AT spectra for different principle quantum numbers ($n$)}

We investigate the AT splitting in cold atoms for Rydberg states with a wide range of principal quantum numbers ($n=35-117$).
Fig.\ref{fig:RydIntATdiffN}(a) shows the AT spectra for different principal quantum numbers, and the corresponding plot from the single-particle model is shown in fig.\ref{fig:RydIntATdiffN} (b). As compared to other studies on AT splitting in cold atoms \cite{2013_Jia,DeSalvo2016, 2021_Jia, 2022_Cao, saakyan2023rydberg} probing Rydberg states with $n < 72$, we investigate the effect of interaction-induced dephasing in AT splitting for highly excited Rydberg states with $n > 100$. The interaction-induced dephasing $\gamma_r$ of the Rydberg state results in the broadening of the spectra for larger principal quantum numbers. We observe a drastic increase in the broadening of the AT spectra for highly excited Rydberg states with principal quantum number $n > 100$. The broadening of the red-detuned peak is much larger compared to that of the blue-detuned peak as shown in Fig.\ref{fig:RydIntATdiffN}  The results of our theoretical modeling (solid lines) agree well with our experimental measurements as shown in Fig.\ref{fig:RydIntATdiffN}.
The increase in the width of the red-detuned peak is proportional to the increase in the interaction-induced dephasing rate as $\gamma_r$ shown in fig.\ref{fig:RydIntlinewidthdiffN}(b). The interaction-induced broadening of the spectra also deteriorates the visibility of AT splitting. All the measurements in Fig.\ref{fig:RydIntATdiffN} are carried out with approximately equal cooling beam intensity of 90 mW/cm$^2$ and the AT splitting almost vanishes for the n=117. The probe beam intensity for different principal quantum numbers is chosen in such a way as to have the same probe Rabi frequency $\Omega_p$. 
The variation of the linewidth of red and blue detuned peaks of the corresponding measurement is shown in Fig.\ref{fig:RydIntlinewidthdiffN}(a). There is a steep increase in the width of the red-detuned peak as compared to the blue-detuned peak when the principal quantum number $n$ of the Rydberg state is increased beyond $n=100$ as shown in Fig.\ref{fig:RydIntlinewidthdiffN}(a). The dephasing rate $\gamma_r$ calculated from our theoretical modeling for Rydberg states with various principal quantum numbers is shown in Fig.\ref{fig:RydIntlinewidthdiffN}(b). The sharp increase in the dephasing rate explains the corresponding increase in the width of the red-detuned peak of the measured AT splitting signal shown in Fig.\ref{fig:RydIntlinewidthdiffN}(a).

\begin{figure*}
\includegraphics[width=0.8\textwidth]{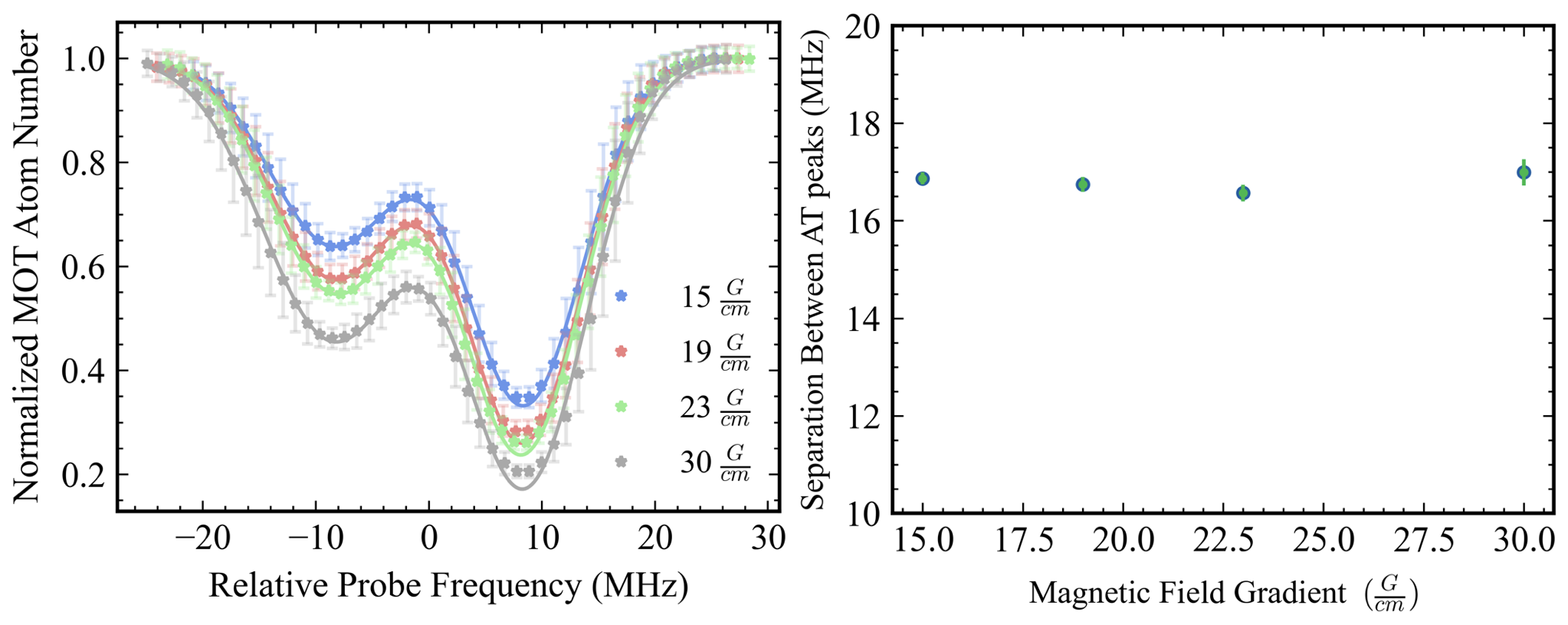}
\caption{\label{fig:35SDiffMagGradAT}The Trap-loss spectra obtained for $\ket{r}=35S_{1/2}$ with different MOT anti-Helmholtz coil magnetic field gradient (Left). The amount A-T splitting vs the magnetic field gradient (Right).}
\end{figure*}
\section{Conclusion}
We have investigated the two-photon excitation to Rydberg states in cold \textsuperscript{87}Rb atoms in a MOT. In a steady-state MOT, the trap-loss spectroscopy method was used to quantify the Rydberg excitation. Although they are no longer constrained by the MOT potential, the atoms excited to the Rydberg state from the cold atomic ensemble will have the same temperature as the cold atoms. This causes the trap-loss dips in the fluorescence measurement of the cold atoms and finally enables these atoms to escape out of the trap.
The two dips in the trap-loss spectra are produced by the AT splitting caused by the strong coupling via the red-detuned cooling beams of the MOT, causing asymmetric AT splitting for the intermediate level. These dips are separated by the generalized Rabi frequency for the cooling transition of the MOT. The strength of the interaction between the atoms excited to the Rydberg state increases with the principal quantum number. In the case of large principal quantum numbers, we also observe a sharp increase in the broadening of the Autler-Townes splitting signals due to high dephasing rate arising from the interatomic interaction between the Rydberg atoms. We perform theoretical modeling and numerical simulation based on the Lindblad Master equation and find good agreement with our experimental observations. By analyzing the resulting AT spectra and explaining the same using theoretical modeling, we aim to contribute to the growing body of knowledge on Rydberg atom interactions and their implications for advanced quantum technologies \cite{2020_Adams, 2024_QT}. 


\begin{acknowledgments}
We acknowledge useful discussions with Weibin Li. S K B acknowledges the funding from the I-HUB Quantum Technology Foundation via the SPIKE Project Grant
No. I-HUB/SPIKE/2023-24/004. RN acknowledges DST-SERB for the Swarnajayanti fellowship (File No. SB/SJF/2020-21/19), MATRICS Grant No. MTR/2022/000454 from SERB. V.S acknowledges the Chanakya fellowship awarded by I-HUB Quantum Technology Foundation, Pune INDIA. We further thank Government of India, National Supercomputing Mission for providing computing resources of ``PARAM Brahma'' at IISER Pune, which is implemented by C-DAC and supported by the Ministry of Electronics and Information Technology and Department of Science and Technology (DST), Government of India, and acknowledge National Mission on Interdisciplinary Cyber-Physical Systems of the Department of Science and Technology, Government of India, through the I-HUB Quantum Technology Foundation, Pune, India.
\end{acknowledgments}

\appendix
\section{Optimization procedure}
\label{aopt}

To determine the interaction-induced dephasing rate $\gamma_r$  and the shift in the cooling detuning $\Delta_c$ we perform an optimization by fitting the AT spectra obtained from our effective single-atom model to the experimental trap-loss spectroscopy results. In this fitting process $\gamma_r$ and $\Delta_c$ are treated as free parameters. To optimize $\gamma_r$ and $\Delta_c$, we minimize a binary cross-entropy loss function, defined as: 
\begin{equation}
    \mathrm{L} = \frac{1}{N} \sum_i y^{exp}_i log(y^{th}_i) + (1 - y^{exp}_i) log(1 - y^{th}_i)
\end{equation}
where $y^{exp}_i$ and $y^{th}_i$ are the experimentally and theoretically obtained values of $1-f_R$ respectively for each probe detuning $\Delta_p$. For optimization, we employ the \textit{scipy.optimize} package, specifically utilizing the COBYLA (Constrained Optimization By Linear Approximations) method to find the optimal values of $\gamma_r$ and $\Delta_c$. 

\vspace{4mm}

\section{A-T Trap loss spectra for different magnetic field gradient of MOT coils}
The magnetic field gradient of MOT anti-Helmholtz coils can be adjusted by changing the current passing through them. The trap is steeper and denser when the magnetic field gradient is larger. Our MOT anti-Helmholtz coils can create a magnetic field gradient of $\sim 10 \frac{G}{cm. A}$. The MOT become steeper and the peak number density will increase when we increase the magnetic field gradient of the coils. The peak atom number in the MOT has varied between 1.1 - 7.0 $\times 10^9 cm^{-3}$ by changing the magnetic field gradient in the range 15-30 $\frac{G}{cm}$. Fig.\ref{fig:35SDiffMagGradAT} shows the A-T trap-loss spectra for $\ket{r}=55S_{1/2}$ by varying the magnetic field gradient. The A-T splitting is almost a constant in the measurement. The non-zero magnetic field, though very small, still can contribute to the broadening of the spectra as all measurements were carried out with MOT coils on. The range of MOT peak density is very small to observe any significant Rydberg interaction-induced dephasing broadening \citep{saakyan2023rydberg,zhang2014autler}. 



\bibliography{apssamp}

\end{document}